# COMPARISON OF CINEPAK, INTEL, MICROSOFT VIDEO AND INDEO CODEC FOR VIDEO COMPRESSION

Suleiman Mustafa[1] and Hannan Xiao[2]

[1]Department of Engineering, National Agency for Science and Engineering Infrastructure, Abuja, Nigeria
[2]School of Computer Science, University of Hertfordshire Infrastructure, Hatfield, UK

***ABSTRACT***

*The file size and picture quality are factors to be considered for streaming, storage and transmitting videos over networks. This work compares Cinepak, Intel, Microsoft Video and Indeo Codec for video compression. The peak signal to noise ratio is used to compare the quality of such video compressed using AVI codecs. The most widely used objective measurement by developers of video processing systems is Peak Signal-to-Noise Ratio (PSNR). Peak Signal to Noise Ration is measured on a logarithmic scale and depends on the mean squared error (MSE) between an original and an impaired image or video, relative to* $(2n\text{-}1)^{2.}$

*Previous research done regarding assessing of video quality has been mainly by the use of subjective methods, and there is still no standard method for objective assessments. Although it has been considered that compression might not be significant in future as storage and transmission capabilities improve, but at low bandwidths compression makes communication possible.*

## 1. INTRODUCTION

High quality video in multimedia applications and wireless communication has generated interest in digital communication services for sharing real-time video audio and data. This is as a result of success and growth in this area. There is an increased demand in providing network portable computers with access to the same services as wired computers [1]. Users of multimedia applications expect a constant level of quality in video. It is necessary to determine the quality of the video images displayed to the viewer when evaluating and comparing video. There has been an increase in the demand for portable computers to provide universal connectivity similar to that of wired networks. The IEEE 802.11 study group was aimed at providing an international standard for WLANs, to satisfy the needs of wireless local area networks [15].

There are two techniques that could have used for evaluation of video quality which are subjective and objective methods. Subjective methods can be difficult to perform and to gain an accurate measure because of subjective factors. Therefore developers prefer objective measures [4]. We decided to use Peak Signal Noise Ratio (PSNR) that is the most common method. It is widely acceptable because it can be calculated easily and measures are repeatable, and for this reason. One problem with the peak-signal to noise ratio is that an original unimpaired video needs to be used to make a comparison. The availability of original image or video signals that are free from distortion or perfect quality affects the type of objective measure that can be performed [10].

## 2. VIDEO COMPRESSION

There are two compression techniques, which are lossy and lossless compressions. Lossless compression is where system statistical redundancy is minimized so that the original signal can be perfectly reconstructed at the receiver only results in modest amount of compression. Lossy compression achieves greater compression except that the decoded signal is not identical to the original. Lossless compression can only achieve a modest amount of compression. Therefore most practical techniques are based on lossy compression, which achieves greater compression, but with loss of decoded signal [2]. The goal of a video compression is to achieve efficient compression whistle minimizing the distortion introduced by the compression process.

### 2.1. Video Compression STANDARDS

This work lies within the MPEG-4 and H 264 compression standards for video compression. MPEG-4 was developed to improve transmission of video over mobile devices. We did research on the previous standards as well to understand how in has been improved for mobile communications. These are MPEG-1, MPEG-2 and MPEG-3, which were all designed for particular applications and bit rate. MPEG-1 was developed for up to 1.5Mbit/sec and widely used for mpg files over the Internet. It is based on CD-ROM video application such as MP3 for digital audio compression. MPEG-2 standard was based on digital television and DVD, developed for 1.5 to 15Mbit/sec. It is an improvement over MPEG-1 for digital broadcast with more efficient compressed interlaced video. MPEG-4 was developed for multimedia and web compression and is an object-based compression [3].

The latest video compression standards MPEG-4 and H.264 standards also known as Advanced Video Coding were important to cover as this work because they deal with multimedia communications and are developed to assist wireless communication networks. MPEG-4 is concerned with providing flexibility while H.264 is concerned with efficiency and reliability [2]. Figure 1 below shows the H.261 encoder system and the revere happens at the decoder. H.261 standard only specifies the decoding for each of the compression options, which allows manufacturers of codec such as those used in this project to differentiate their product [6].

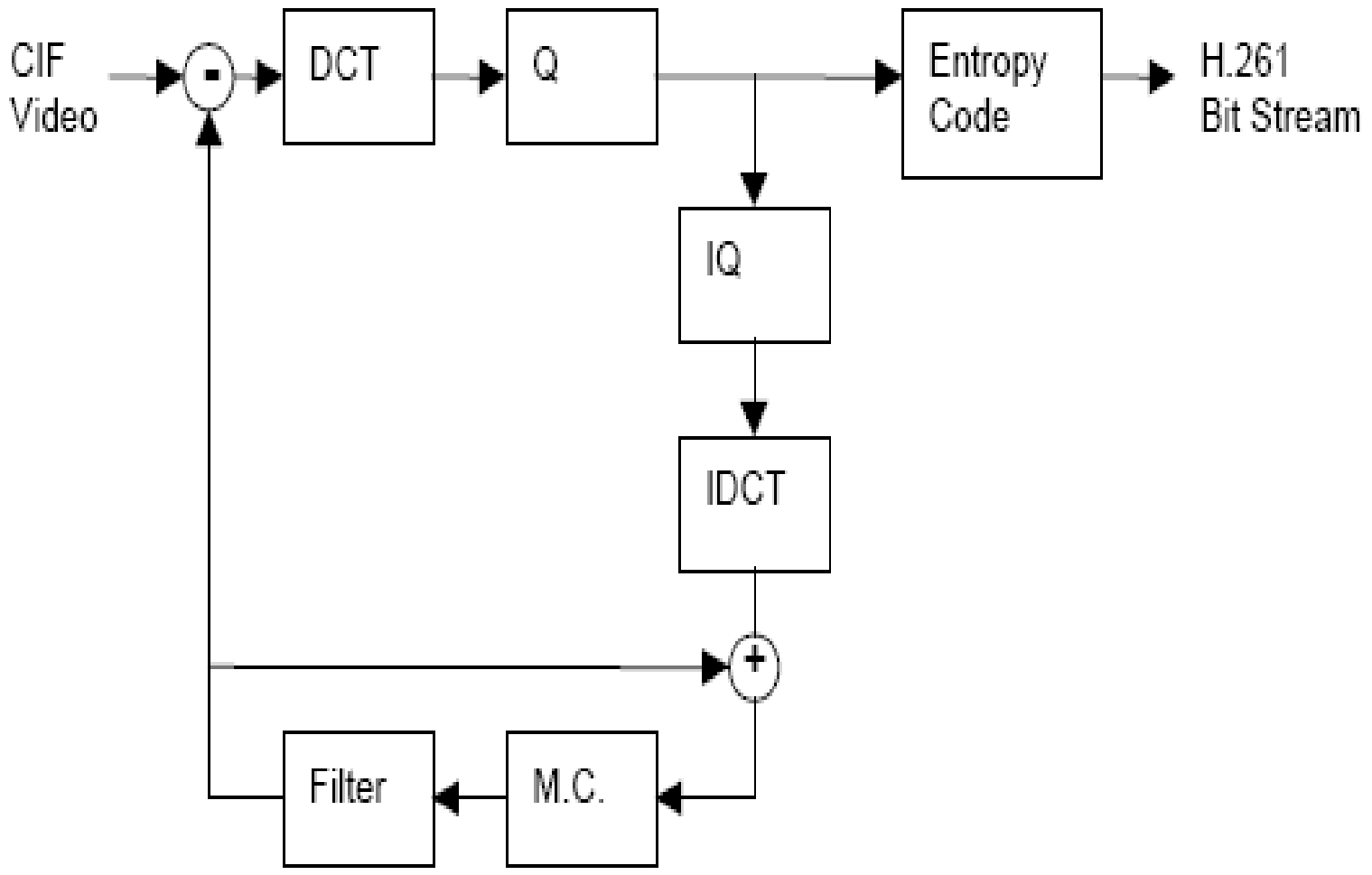

Figure 1 H.261 Encoder

*Source: (Array Microsystems, Inc Video Compression white paper, 1997)*

## 2.1. Objective Measurement of Video Quality

Quality metrics that can predict image and video quality automatically is the goal of objective video quality assessment research. Common methods are mean square error (MSE) and peak signal-to-noise ratio (PNSR) which make use of original reference [7].

For future development in objective measurement of video quality, peak signal-to-noise ratio could be replaced with a metric that closely matches the behavior of the human visual system. Color and motion plays a key role in defining the way we perceive video quality [4]. Calculating distortion between videos may be unrealistic because the distortion occurs where the human eye is less sensitive. Therefore methods that match closely to the subjective perception of the eye would help produce more accurate and realistic results. With more time available the project could be improved upon by using subjective methods to verify the objective measurements. This would involve users to get judgments on which videos are of better quality.

The ITU-T video Quality Experts Group (VQEG) aims to compare and run test on potential models of evaluation objectively. There is still no standard accurate system for digitally decoded video but the aim is to have an automatic system to solve the problem of accurate measure. The new generation of advanced video compression technologies, including MPEG-4 Part 10 (also known as H.264/MPEG-4 AVC) and Windows Media Player 9, are key to the distribution of broadcast-quality video over IP networks. Advanced video compression reduces bandwidth requirements for high-quality video by about half compared to existing MPEG-2 technology. The ability to implement the new advanced video compression standards has important benefits for service providers [18].

From the work of the ITU study group VQEG (Visual Quality Experts' Group); several sophisticated quality measures were compared with PSNR. The results showed that for MPEG-2 coding at bit rates equivalent to 0.8Mbits/sec or more, none of the measures was significantly more accurate than PSNR [16]. It can therefore be safely concluded that over some period of time PSNR may be used as a benchmark for measuring video quality.

## 2.2. Codec Video

Coded video is produced at a variable or constant rate by a video encoder, and the average bit rate and the bit rate variation are the important parameters. Distortion is introduced because the original video signals are not similar to decoded video sequence. For quality of service, delay depends on the method used for transmission such as broadcast, streaming, playback or video conferencing [4]. Coded video has very low tolerance to delay unlike data. Dropped video information cannot be retransmitted therefore a method that enables error control is used for compressed video data [5]. The choice of what codec to use can be difficult because manufacturers present their capabilities in different ways that best suit their product. The majority of commercial ones seem to be aimed for streaming or storage applications [2].

The increase demand to incorporate video data into telecommunication services, the corporate environment, the entertainment industry and the home has made video technology a necessity. However it is still a problem that digital video data rates are very large and require a lot of bandwidth. There are various forms in which uncompressed videos are encoded and decoded [2]. For this project the codec are all AVI video codec and there are two formats for coding AVI video files used. The video data in an AVI file can be formatted and compressed in a variety of ways [24].

## 3. RELATED WORK

Early researches aimed at comparison of video codec have used both subjective comparison with convenient visualization and detailed objective comparisons. Series of subjective comparison of popular codecs have been carried out. In one of such comparisons, all codecs were tested in a 2-pass setup using the settings suggested by the developers. Various stages of testing were carried out using a wide range of codec. The best ones according to the research were Ateme, x264, XviD and DivX 6.1 in that order [22].
The main aim of comparing various codec is to investigate which co mpression will produce a video of the smallest possible file size that is most close to the original video. Most objective research has made use of peak signal-to-noise ratio as a video quality metric [21]. MUS MPEG-4 SP/ASP Codec comparison for instance tested to compare different versions of MPEG-4 codec. Tracing the evolution of the DivX codec was another goal of the testing which showed that the latest DivX had the advantage with better quality of compression performed by this codec over XviD [20].

Comparison of the Advanced Video Coding (AVC/H.264) standard by the VideoLAN x264 project, the VP8 codec provided by Google/WebM project and HEVC TMuC v0.5 has also been carried out in another research. Videos content for streaming were tested for mobile devices. For the same video distortion in terms of PSNR and structural similarities, revealed that on average HEVC TMuC v0.5 provides 46% bitrate reduction in comparison to AVC/H.264, whuch in turn provides 21% bitrate reduction compared to Google/WebM VP8 [23].

## 4. EXPERIMENT DESIGN

This work compares Cinepak, Intel, Microsoft Video and Indeo Codec for video compression. Performing experiments to measure the quality of video compressed using Radius Cinepak, Intel Indeo R3.2, Intel Indeo 4.5, Microsoft video 1 and Indeo Video codec did the research. The method used for comparing the videos is peak-signal-to noise ratio. It is calculated by using the mean square error (MSE) between an original and a compressed video frame relative to the square of the highest possible signal value in the image $(2^n-1)^2$ [4].

### 4.1. Methodology

The method used for the measurement of the video quality for this work is peak signal-to-noise ratio (PSNR). It is calculated by using the mean square error (MSE) between an original and a compressed video frame, relative to the square of the highest possible signal value in the image $(2n-1)^2$ [4]. Because of the two-dimensional matrix nature of a digital picture, SNR for an image can be considered a matrix based quality parameter which is the peak signal to noise ratio in of a digital video. The word peak refers to the maximum value of a pixel [7]. The unit of measure is given in decibel units (dB). It relies on the pixel luminance and chrominance values of the input and output video frames. It does not include any subjective human intervention in the quality assessment [5].

As mentioned earlier in the project introduction, PSNR measures the difference between two images and a reference and compressed image is used. For this project the reference or unimpaired videos are the original uncompressed ones and the compressed videos are those converted using the codec. PSNR analysis uses a standard mathematical model to measure an objective difference between two images. It is commonly used in the development and analysis of compression algorithms, and for comparing visual quality between different compression systems. PSNR = 10*log10 (255^2/MSE)

*Pseudo code for each pixel:*

*{*

*Difference = Pixel from Image A – Pixel from Image B*

*SummedError = SummedError + Difference * Difference*

*}*

*MeanSquaredError = SummedError / Number of Pixels*

*RMSE = sqrtt (MeanSquaredError)*

*PSNR = 20*log10 (255 / RMSE)*

Where original video uses 8 bits per pixel and therefore the peak is 255.

*Source: http://www.cineform.com/technology/HDQualityAnalysis10bit/HDmethodology10bit.htm*

There are three values produced for each frame Y, U and V which represent the luminance and chrominance. A frame consists of the three rectangular arrays of integer valued-samples. Y is called luma and represents brightness, while U and V are the two Chroma sometimes represented as Cb and Cr respectively. U (Cb) represents the extent to which the colour deviates from gray to blue, while V (Cr) represents the extent to which the colour deviates from gray to red [8].

## 4.2. Choosing a Video Quality Metric

The Video Quality Studio 0.4 RC3 software used measures the luminance and chrominance of each frame of the video files. There are two sections for selecting video files from the hard disk, one for selecting the reference video and the other for the compressed/impaired video. The output is produced on an excel file which shows the Y, U and V components of each frame. Videos coding often uses a colour representation having three components of a tristimulus colour representation for the spiral area represented in the image [8].

Attempts made in order to design a method for objective picture quality measure that improves upon PSNR. This method is based on the concept of Just Noticeable Differences (JND) and equipment that perform PSNR and JND-based measures are available. Specific single–ended measurements of impairments introduced by compression have received some attention. This method makes use of decoded video and information from the bit stream as well [16]. Objective criteria produce accurate and repeatable results but there are yet no objective measurement systems that completely reproduce the subjective experience when watching a video [4].

## 4.3. Selecting Video files

The video files had to be in AVI format for the Video Quality Studio 0.4 RC3 software to use. AVI (Audio Video Interleave) as defined by Microsoft are basically a number of still images called frames that are combined sequentially in one file. When the file is opened with a media player it moves through the AVI file and displays each consecutive image in the same way that movie film rolling through a projector displays a movie playing [11]. To notice the effect of the compression we decided to use video files of different properties such as length of video, content of video, size of file etc. The videos use range from those that are just a few seconds to those that are up to a minute long, fast moving pictures and slow moving pictures and other characteristics. All the videos used for the experiment have a frame rate = 23 frames/second, video sample size = 24 bits, and audio sample size = 16 bits. The original uncompressed reference video was recorded with a 10.1 megapixel camera. The videos were converted to five different AVI formats using the following codec: Radius Cinepak, Intel Indeo R3.2, Intel Indeo 4.5, Indeo Video 5.10 and Microsoft Video 1. The table below shows the list of video files used:

Table 1 List of Video files and their properties.

| **File Name** | **Codec Type** | **Data rate (kbps)** | **Size (MB)** |
|---|---|---|---|
| Frontalis- Video Clip[Cinepak] | Cinepak codec by Radius | 322 | 15.8 |
| Frontalis- Video Clip[II 4.5] | Intel Indeo video R3.2 | 299 | 10.2 |
| Frontalis- Video Clip[II r3.2] | Intel Indeo video 4.5 | 214 | 14.3 |
| Frontalis- Video Clip[Indeo video] | Indeo video 5.10 | 261 | 12.5 |
| Frontalis- Video Clip[Microsoft] | Microsoft Video 1 | 861 | 41.3 |
| Frontalis- Video Clip[reference] | (No codec) | 1411 | 101 |
| Infraspinatus- Video Clip[Cinepak] | Cinepak codec by Radius | 344 | 16.9 |
| Infraspinatus- Video Clip[ II 4.5] | Intel Indeo video R3.2 | 334 | 16.4 |
| Infraspinatus- Video Clip[II r3.2] | Intel Indeo video 4.5 | 230 | 11.3 |
| Infraspinatus- Video Clip[Indeo video] | Indeo video 5.10 | 189 | 13.7 |
| Infraspinatus- Video Clip[Microsoft] | Microsoft Video 1 | 883 | 43.4 |
| Infraspinatus- Video Clip[reference] | (No codec) | 1435 | 104 |
| Intro Question [Cinepak] | Cinepak codec by Radius | 495 | 24.3 |
| Intro Question [II 4.5] | Intel Indeo video R3.2 | 481 | 23.6 |
| Intro Question [II r3.2] | Intel Indeo video 4.5 | 331 | 16.3 |
| Intro Question [Indeo video] | Indeo video 5.10 | 272 | 19.7 |
| Intro Question [Microsoft] | Microsoft Video 1 | 1271 | 62.5 |
| Intro Question [reference] | (No codec) | 2066 | 150 |
| Quiz 02[Cinepak] | Cinepak codec by Radius | 155 | 7.6 |
| Quiz 02[II 4.5] | Intel Indeo video 4.5 | 241 | 8.2 |
| Quiz 02[II r3.2] | Intel Indeo video R3.2 | 136 | 9.1 |
| Quiz 02[Indeo video] | Indeo video 5.10 | 177 | 8.5 |
| Quiz 02[Microsoft] | Microsoft Video 1 | 757 | 36.3 |
| Quiz 02[reference] | (No codec) | 1041 | 74.5 |
| Race[Cinepak] | Cinepak codec by Radius | 322 | 15.8 |
| Race[II 4.5] | Intel Indeo video 4.5 | 299 | 10.2 |
| Race[II r3.2] | Intel Indeo video R3.2 | 214 | 14.3 |
| Race[Indeo video] | Indeo video 5.10 | 261 | 12.5 |
| Race[Microsoft] | Microsoft Video 1 | 861 | 41.3 |
| Race[reference] | (No codec) | 1411 | 101 |
| Alien Eye [Cinepak] | Cinepak codec by Radius | 837 | 41.1 |
| Alien Eye [II 4.5] | Intel Indeo video 4.5 | 777 | 26.5 |
| Alien Eye [II r3.2] | Intel Indeo video R3.2 | 556 | 37.2 |
| Alien Eye [Indeo video] | Indeo video 5.10 | 601 | 32.5 |
| Alien Eye [Microsoft] | Microsoft Video 1 | 3669 | 107.4 |
| Alien Eye [reference] | (No codec) | 2176 | 262.6 |

## 4.4. Choosing Codec

The AVI MPEG video converter software to convert other forms of video files to AVI format. Another advantage of the software was that we could convert to different forms of AVI videos that use the different codec tested. The codec that it uses are Cinepak, Intel, Microsoft Video, Indeo, Xvid and Divx codec, but we could not get all the videos to convert with Xvid and Divx. The rest of the codec, which are Cinepak, Intel, Microsoft Video and Indeo all successfully, convert the videos and were then used.

The manufactures tend to present their codec in ways that favor their product (I.E.G Richardson, 2002), but by studying them we gained a better understanding of how they work. There are three common color-encoding schemes used in broadcast television systems around the world. Most European countries and many other nations around the world use Phase Alternating Line (PAL). It was introduced first in Britain and Germany in 1967. The United States and Japan use another standard the NTSC (National Television System Committee). France and a few other nations use the SECAM (Sequential Couleur Avec Memoire or Sequential Color with Memory) standard. Y refers to the luminance, a weighted sum of the red, green, and blue components. The human visual system is most sensitive to the luminance component of an image. Analog video systems such as NTSC, PAL, and SECAM transmit color video signals as a luminance (Y) signal and two color difference or chrominance signals [11]. Below is some brief information about the codec used for compressing the videos in this experiment

### 4.4.1. Cinepak

The codec is a vector quantization based on image compression and frame with adaptive vector density and each frame is segmented into 4x4 pixel blocks and each block is coded using 1 or 4 vectors. Intel Indeo also uses vector quantization based image compression. Rather than bit rate versus quality performance, radius Cinepak comes from computational simplicity at the decoder. This codec compressed most videos successfully at a quick rate [12].

### 3.4.2. Intel Indeo R3.2 and 4.5

Indeo was originally known as Real Time Video when Intel developed it in the 1980's. Indeo is very similar to the Cinepak codec. It is well suited to CD-ROM, has fairly high compression times, and plays back on a wide variety of machines. The recommended key frame interval for Indeo is 4, regardless of the frame rate. Both codec compressed all videos successfully at a fast rate [13].

### 4.4.3. Indeo Video 5.10

This is a new wavelet compression algorithm that greatly improves visual quality. It has been optimized to provide the best possible performance for Indeo on fast systems especially with newer scalability features. This codec compressed only some videos and took a longer time to do compare to all the other codec. Those that could not be compressed were not used for the experiments. For this same reason the number of videos originally planned for were reduced [14].

#### 4.4.4. Microsoft Video

The Microsoft Video-1 codec has a simpler algorithm compared with other modern video compression methods. The algorithm operates on 4x4 blocks of pixels, which implies that the source data to be compressed must be divisible by 4 in both its width and height. Just like decoding a Microsoft BMP image, decoding a frame of Video-1 data is a bottom-to-top operation [17]. The codec did not compress as fast as Cinepak or the Intel Indeo codec but was faster than Indeo Video. It also compressed only some of the videos, and those, which could not be compressed, were not used for the experiment.

## 5. EXPERIMENT RESULTS

The values Y, U and V are luminance and chrominance of the video as discussed in the experiment methodology earlier. Y is the luma and represents brightness, while U and V are the two chroma represented at times as Cb and Cr respectively. U (Cb) represents the extent to which the color deviates from gray to blue, while V (Cr) represents the extent to which the color deviates from gray to red [8]. YUV is the color space used in the European PAL broadcast television standard. U is very similar to the difference between the blue and yellow components of a color image. V is very similar to the difference between the red and green components of a color image. There is evidence that the human visual system processes color information into something like a luminance channel, a blue – yellow channel, and a red - green channel. For example, while we perceive blue-green hues, we never perceive a hue that is simultaneously blue and yellow. This may be why the YUV color space of PAL is so useful [11].

The unit for peak signal-to-noise ratio is decibel units (dB) and a higher number indicates better quality video. PSNR = 10*log10 (255^2/MSE), the original videos use 8 bits per Pixel, therefore the peak = 255. Peak signal-to-noise ratio values greater than 35 dB are considered to be good quality. As this project deals with colored videos the average values for the Y, U and V components are indicated. The picture resolution and frame rate need to be indicated because bit rate is directly proportional to the number of pixel per frame and the number of frames coded per second. The total number of frames for each video, video resolution and the length are shown below. A frame rate of 23 frames per second was adopted for all the experiments carried out. The results shown below are the average peak signal-to-noise ratio values for the total number of frames and the standard deviation for the peak signal-to-noise ratio values of the total number of frames in each video. The tables below show the results for average and standard deviation for each video and codec.

Table 2 **Infraspinatus Video**: – Frames = 1215 Frames, Length = 50 seconds, Resolution = 192 x 144

| | Radius Cinepak | Intel Indeo video R3.2 | Intel Indeo video 4.5 | Indeo video 5.10 | Microsoft Video 1 |
|---|---|---|---|---|---|
| Average | Y=13.7345<br>U=25.0233<br>V=23.3655 | Y=6.53445<br>U=20.2398<br>V=21.6978 | Y = 39.1169<br>U = 42.3003<br>V = 42.0153 | Y = 42.5232<br>U = 43.3078<br>V = 43.305 | Y = 35.8655<br>U = 46.5737<br>V = 45.4756 |
| Standard Deviation | Y=0.428589<br>U=0.367764<br>V=0.407782 | Y=0.072323<br>U =0.33221<br>V=0.325263 | Y=0.594707<br>U=0.701137<br>V = 0.59316 | Y=0.997392<br>U=0.793095<br>V=0.706637 | Y = 0.234032<br>U = 0.448261<br>V = 0.442202 |

Table 3 **Frontalis Video**:– Frames = 1178 Frames, Length = 49 seconds, Resolution = 192 x 144

| | Radius Cinepak | Intel Indeo video R3.2 | Intel Indeo video 4.5 | Indeo video 5.10 | Microsoft Video 1 |
|---|---|---|---|---|---|
| Average | Y=12.4925<br>U =21.21<br>V=17.9829 | Y = 6.17462<br>U = 19.7556<br>V = 20.3806 | Y = 43.0715<br>U = 43.9592<br>V = 43.5472 | Y = 43.1586<br>U = 43.8729<br>V = 43.3251 | Y = 37.0314<br>U = 45.4159<br>V = 44.7032 |
| Standard Deviation | Y=1.14014<br>U=0.715839<br>V=0.708884 | Y=0.162804<br>U =0.3165<br>V=0.313484 | Y=1.22018<br>U=1.14076<br>V=0.715536 | Y=1.051<br>U=0.791641<br>V=0.615129 | Y=0.232984<br>U=0.218801<br>V=0.185936 |

Table 4 **Intro Question Video**:- Frames = 1097 Frames, Length = 45 seconds, Resolution = 400 x 448

| | Radius Cinepak | Intel Indeo video R3.2 | Intel Indeo video 4.5 | Indeo video 5.10 | Microsoft Video 1 |
|---|---|---|---|---|---|
| Average | Y = 15.7287<br>U = 24.7749<br>V = 40.529 | Y = 13.118<br>U = 13.5661<br>V = 29.7467 | Y = 34.9369<br>U = 29.6101<br>V = 43.7315 | Y = 49.7683<br>U = 29.5921<br>V = 43.9668 | Y =46.7505<br>U =30.7454<br>V =45.3311 |
| Standard Deviation | Y=0.696676<br>U=0.626488<br>V=0.628696 | Y=0.279379<br>U=0.0473019<br>V=0.0485187 | Y=0.544054<br>U = 0.81499<br>V=0.563606 | Y= 0.67752<br>U=0.810964<br>V=0.613135 | Y=0.927657<br>U=0.843353<br>V = 0.65692 |

Table 5 **Quiz 02 Video**: - Frames = 1178 Frames, Length = 49 seconds, Resolution = 192 x 144

| | Cinepak Radius | Intel Indeo video R3.2 | Intel Indeo video 4.5 | Indeo video 5.10 | Microsoft Video 1 |
|---|---|---|---|---|---|
| Average | Y =16.8062<br>U =27.8529<br>V =41.8063 | Y =14.0409<br>U =13.6811<br>V =29.9228 | Y = 32.672<br>U = 36.009<br>V = 46.6161 | Y = 47.8617<br>U = 35.9763<br>V = 46.9077 | Y = 42.7775<br>U = 39.2559<br>V = 55.2206 |
| Standard Deviation | Y =0.26265<br>U=0.230529<br>V=0.139003 | Y=0.103465<br>U=0.0179045<br>V=0.0202713 | Y=0.488761<br>U=0.271639<br>V=0.102116 | Y=0.354362<br>U=0.266689<br>V= 0.18454 | Y = 0.24554<br>U=0.0500527<br>V = 0.272689 |

Table 6 **Race Video**:- Frames =392, Length =16 seconds, Resolution = 352 x 240

| | Radius Cinepak | Intel Indeo video R3.2 | Intel Indeo video 4.5 | Indeo video 5.10 | Microsoft Video 1 |
|---|---|---|---|---|---|
| Average | Y=16.2319<br>U=30.1547<br>V=33.2701 | Y=10.9564<br>U=29.3925<br>V=30.6536 | Y=44.3539<br>U=44.2083<br>V=47.3081 | Y = 45.8188<br>U = 44.4679<br>V = 47.5591 | Y = 38.9927<br>U = 46.9736<br>V = 47.7328 |
| Standard Deviation | Y=1.89294<br>U=4.91656<br>V=4.56857 | Y=1.35758<br>U=1.49487<br>V=1.08814 | Y=2.16613<br>U=1.32188<br>V = 1.3906 | Y= 1.45584<br>U = 1.14082<br>V=0.952739 | Y = 0.349535<br>U = 0.311646<br>V = 0.252711 |

Table 7 **Alien Eye Video**:- Frames =382, Length =15 seconds, Resolution = 352 x 240

| | Radius Cinepak | Intel Indeo video R3.2 | Intel Indeo video 4.5 | Indeo video 5.10 | Microsoft Video 1 |
|---|---|---|---|---|---|
| Average | Y=17.0744<br>U=37.0833<br>V=38.4051 | Y=11.8769<br>U=37.4427<br>V=37.5588 | Y = 41.2606<br>U = 46.9533<br>V = 48.6508 | Y = 42.4673<br>U = 46.8702<br>V = 48.4842 | Y = 36.5074<br>U = 46.2448<br>V = 45.6937 |
| Standard Deviation | Y=2.77705<br>U=2.17072<br>V=2.36064 | Y=2.76043<br>U =2.2202<br>V =2.3764 | Y=1.64291<br>U=1.16645<br>V=0.976556 | Y=1.19747<br>U =1.08711<br>V=0.922248 | Y = 0.492858<br>U = 0.216027<br>V = 0.268167 |

### 5.1. Analysis and Interpretation of Results

The codec used successfully compressed the files to less than 50% the original file sizes and in most cases up to 70% less in size to the reference video. In terms of file video file sizes from smallest to largest codec the results were Intel Indeo video R3.2, Indeo video 5.10, Intel Indeo video 4.5, Radius Cinepak and then Microsoft Video 1. Although there was not much significant difference in the files sizes of the compressed videos except for Microsoft Video 1 codec, which was much larger in size. In terms of the PSNR, the values for Y,U and V from highest to lowest codec were Indeo video 5.10, Intel Indeo video 4.5, Microsoft Video 1, Radius Cinepak and finally Intel Indeo video R3.2.

## 6. CONCLUSIONS

The results of the experiment show that the codec used produced variable results but were consistent over the sets of videos tested. From the experiment result Indeo video 5.10 had the highest PSNR values and the second smallest file sizes and therefore the codec is recommended best for AVI file format.

PSNR does not usually correlate well with the other subjective quality measurements all the time. This is because changes in pixel level particularly in the U and V chrominance are not always detected by the human visual system and can be ignored. Therefore another solution to give more realistic results would be to calculate the luma (Y). This is because luma is the best channel for detecting errors that the human eye is most likely to perceive. Also, by analyzing the luma channel directly. We avoid any PSNR error that may occur in YUV to RGB conversion; therefore we get a more accurate measure of the codec's quality performance [19].

An alternative to YUV is the RGB (red/green/blue) color space, where three numbers for the proportions of the three primary colors red, green and blue represent each pixel. The human eye is more sensitive to brightness than color, which makes RGB not very effective [4]. Each color component picture is partitioned into 8 x 8 pixel blocks of picture samples. Instead of coding each block separately, H.261 groups 4 Y blocks, 1 U block and 1 V block together. MPEG-4 compression algorithm was designed to address the need for higher picture and increased system flexibility. Similar to H.261 as only the YUV color components separation is allowed by the standard, but unlike H.261 the frame size is not fixed. The numbers indicates the brightness or luminance, and a larger number means a brighter sample. For a sample represented by n bits, 0 would represent black and (2^n-1) represents white [4]. 8 bits per sample is usually used to represent luminance for videos such as those used for the experiment.

## ACKNOWLEDGEMENT

We would like to acknowledge the University of Hertfordshire Multimedia Laboratory and the staff for their assistance and faculties used for conducting this research.

## AUTHORS

Suleiman Mustafa presently works with National Agency for Science and Engineering Infrastructure (NASENI), Abuja Nigeria as a Program Analyst since 2010. He holds a Master's degree in Multimedia Technology (2006) and a Bachelor's degree in Computer Science (2004) both from the University of Hertfordshire, UK. He is a member of the Nigeria Computer Society (NCS) and Computer Professionals of Nigeria (CPN).

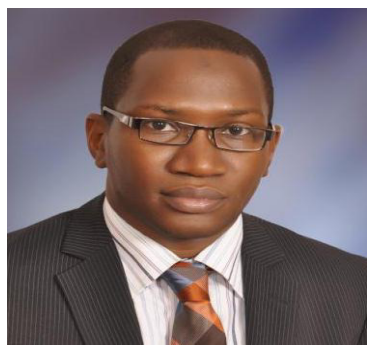

Dr Hannan Xiao received her PhD from the Department of Electrical & Computer Engineering, the National University of Singapore in 2003, and B.Eng and M.Eng degrees from the Department of Electronics & Information System Engineering, the Huazhong University of Science & Technology, China. Dr Xiao has been with the University of Hertfordshire as a lecturer/senior lecturer since October 2003.

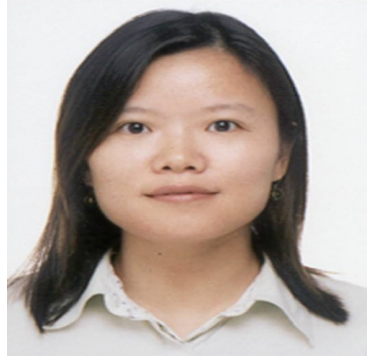